\documentclass[aip,sd,amsmath,amssymb,reprint]{revtex4-1}

\usepackage{graphicx}
\usepackage{dcolumn}
\usepackage{bm}
\usepackage{epstopdf}
\usepackage{xcolor}
\usepackage{natbib}
\usepackage{subfigure} 
\begin{document}

\title{Beyond the ponderomotive limit: direct laser acceleration of relativistic electrons in sub-critical plasmas}


\author{A. V. Arefiev}

\author{V. N. Khudik}
\affiliation{Institute for Fusion Studies, The University of Texas, Austin, Texas 78712, USA}

\author{A. P. L. Robinson}
\affiliation{Central Laser Facility, STFC Rutherford-Appleton Laboratory, Didcot OX11 0QX, United Kingdom}

\author{G. Shvets}
\affiliation{Institute for Fusion Studies, The University of Texas, Austin, Texas 78712, USA}

\author{L. Willingale}
\affiliation{University of Michigan, 2200 Bonisteel Boulevard, Ann Arbor, Michigan 48109, USA}

\author{M. Schollmeier}
\affiliation{Sandia National Laboratories, Albuquerque, New Mexico 87185, USA}

\date{\today}

\begin{abstract}
We examine a regime in which a linearly-polarized laser pulse with relativistic intensity irradiates a sub-critical plasma for much longer than the characteristic electron response time. A steady-state channel is formed in the plasma in this case with quasi-static transverse and longitudinal electric fields. These relatively weak fields significantly alter the electron dynamics. The longitudinal electric field reduces the longitudinal dephasing between the electron and the wave, leading to an enhancement of the electron energy gain from the pulse. The energy gain in this regime is ultimately limited by the superluminosity of the wave fronts induced by the plasma in the channel. The transverse electric field alters the oscillations of the transverse electron velocity, allowing it to remain anti-parallel to laser electric field and leading to a significant energy gain. The energy enhancement is accompanied by development of significant oscillations perpendicular to the plane of the driven motion, making trajectories of energetic electrons three-dimensional. Proper electron injection into the laser beam can further boost the electron energy gain. 
\end{abstract}

\maketitle

\section{Introduction}

The rapid development and improvement of ultra-intense laser pulses have opened new areas of physics for fundamental research and have also enabled novel technological applications. Relativistically intense ($I > 10^{18}$ W/cm$^2$) laser pulses readily ionize matter converting it into a plasma. The interaction with the plasma electrons is the primary channel for the energy transfer from the laser pulse, providing the basis for a wide range of phenomena and applications. Specifically, generation of copious relativistic electrons is the key to x-ray~\cite{Park2006,Kneip2008} and secondary particle sources, such as energetic ions~\cite{Schollmeier2015}, neutrons~\cite{Pomerantz2014}, and positrons~\cite{Chen2015}. It is therefore critical to understand what controls the generation of relativistic electrons in relevant regimes of laser-plasma interactions.

It is well recognized that the regime in which a relativistically intense laser pulse irradiates a sub-critical plasma is optimal for generating relativistic electrons. The pulse can propagate through such a plasma, which enables an extended interaction length with the electrons. In experiments aimed at x-ray generation, this regime is deliberately achieved by using an expanding gas jet~\cite{Kneip2008}. This regime can also naturally arise in experiments with solid density targets irradiated by a powerful laser pulse due to the presence of a prepulse~\cite{Schollmeier2015}. The prepulse often delivers a considerable amount of energy, causing the front of the target to expand and form an extended subcritical preplasma prior to the arrival of the main pulse. 

The important role played by the interaction of the main pulse with the preplasma in generating an energetic electron population in experiments with initially solid-density targets was however not immediately recognized. This is in part due to the fact that in some setups the role of the prepulse can also be detrimental. It is the case in experiments aimed at ion acceleration, where the pre-expansion at the surface of a thick bulk target caused by the prepulse significantly reduces the effectiveness of ion acceleration~\cite{Kaluza2004,Fuchs2007}. Recently, the focus of the ion acceleration research has markedly shifted towards those regimes in which the target becomes transparent to the main relativistically intense pulse and the electron acceleration and heating can be fully utilized~\cite{Powell2015, Palaniyappan2015}. Such a regime is achieved by employing an initially ultra-thin target that significantly expands during the pre-pulse and forms a relativistically transparent plasma extending many wavelengths along the direction of the laser beam propagation. The same regime has also been used to generate energetic electrons that are subsequently converted into a short neutron beam~\cite{Pomerantz2014}.



How energetic electrons are generated strongly depends on the duration of the main laser pulse. In most setups aimed at producing secondary particle sources, the laser pulse irradiates the plasma over a time period that is longer than the characteristic electron response time. This regime is the opposite to the regime used for the wakefield acceleration~\cite{Esarey2009}. As a consequence, the laser pulse establishes a quasi-steady-state structure in the plasma that slowly evolves on an ion time scale~\cite{Mangles2005,Willingale2011,Willingale2013}. By expelling some of the electrons radially, the laser creates a positively charged elongated channel in the plasma with quasi-static transverse and longitudinal electric fields. New electrons are continuously injected into the channel, typically through the channel opening~\cite{Robinson2013,Krygier2014}, and then get accelerated and pushed forward by the laser pulse in the presence of the quasi-static fields. 

\begin{figure*}[htb]
	\centering
	\includegraphics[width=1\linewidth]{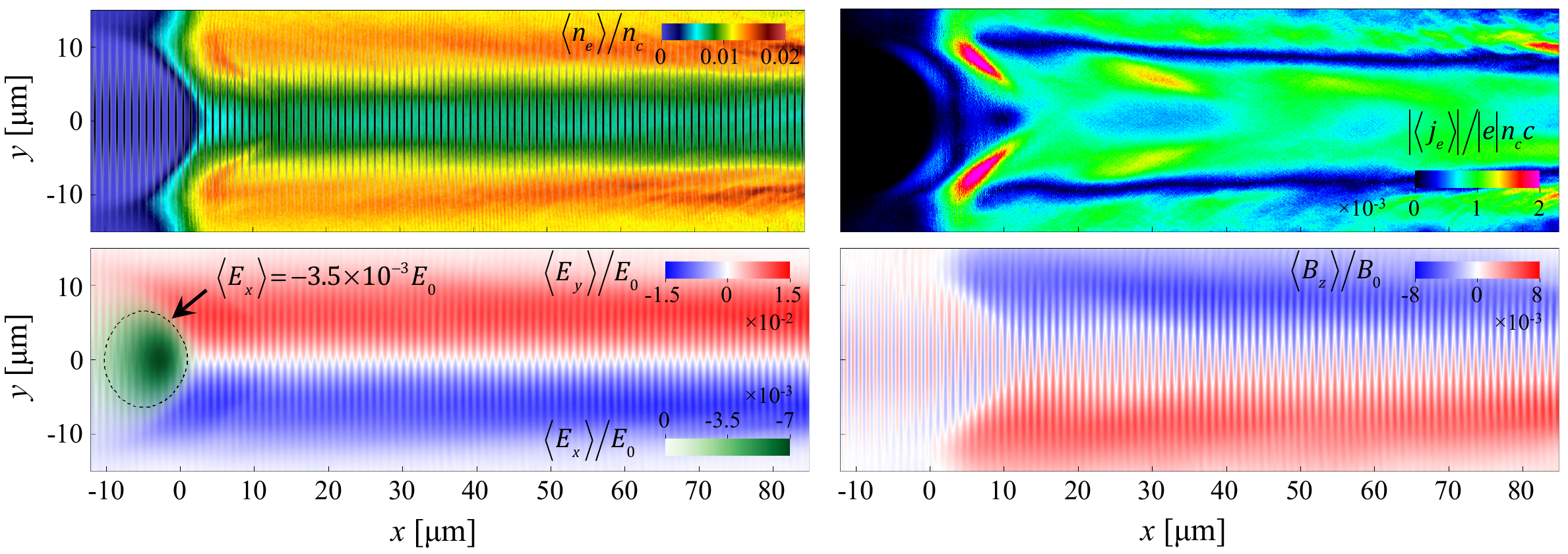}
	\caption{Steady-state channel produced in a sub-critical plasma by a long laser pulse. The panels show snapshots at 1 ps of electron density (upper left), total electron current density (upper right), transverse and longitudinal electric fields (lower left), and transverse magnetic field (lower right) averaged over ten laser periods. The upper left panel also shows instantaneous total electric field.} \label{fig:2D_example}
\end{figure*}

The described regime is often broadly referred to as the direct laser acceleration regime. In terms of applications, it is critical to know what controls the electron energy gain. Early work on the topic~\cite{Pukhov1999} indicated that the transverse static electric field can be beneficial for enhancing the electron energy gain beyond what is expected from a single electron irradiated by a plane wave in a vacuum. Recently, there has been a renewed interest in the direct laser acceleration of electrons, as experimental groups shift their focus to regimes of relativistic transparency  and also try to optimize the preplasma conditions using multiple laser pulses~\cite{Peebles2016}. Recent simulation results have also demonstrated that the direct laser acceleration can be important in the context of the laser wakefield acceleration~\cite{Shaw2014,Zhang2015}. 

In this paper, we examine the role played by transverse and longitudinal quasi-static electric fields present in a plasma channel in enhancing the electron energy gain from the laser pulse. We also address the role of electron injection into the laser beam and the limitations imposed by the super-luminosity of the laser field that is induced by the channel. This paper is based on a body of work performed by us over the last couple of years~\cite{Arefiev2012,Robinson2013,Arefiev2014,Arefiev2015,Robinson2015,Arefiev2015b} and is designed to serve in part as an overview of some novel aspects of the direct laser acceleration. Here we focus on illustrating the key qualitative concepts and phenomena, while providing references to those publications where one can find more detailed technical analysis. 


\section{Steady-state channel} \label{Sec-channel}

The nature of the laser-plasma interaction strongly depends on the amplitude of the irradiating laser pulse and on the electron density of the irradiated plasma. It is convenient to use a dimensionless parameter
\begin{equation}
a_0 \equiv \frac{|e| E_0}{m_e \omega c}
\end{equation}
to quantify the impact of a laser pulse with electric field amplitude $E_0$ and frequency $\omega$ on electron motion. Here $c$ is the speed of light and $e$ and $m_e$ are the electron charge and mass. The parameter $a_0$ is often referred to as the normalized laser amplitude. It is roughly the ratio of the transverse electron momentum induced by the oscillating laser electric field to $m_e c$. Therefore, a laser pulse with a normalized amplitude of $a_0 \geq 1$ would induce relativistic electron motion.

The electron density in the plasma determines whether the laser pulse can propagate into the plasma and accelerate plasma electrons. The cut-off for a pulse with $a_0 \ll 1$ occurs at a critical density,
\begin{equation}
n_c \equiv   \frac{m_e \omega^2}{4 \pi e^2},
\end{equation}
for which the electron plasma frequency $\omega_{pe} = \sqrt{4 \pi n_e e^2 / m_e}$ becomes equal to the frequency of the laser pulse. At laser amplitudes $a_0 \geq 1$, the plasma can become relativistically transparent at electron densities exceeding the critical density $n_c$. The adjusted critical density in this case depends on the amplitude of the irradiating laser pulse, because the effect is caused by the relativistic motion of electrons in the strong field of the laser. 

The optimal regime for generating copious relativistic electrons is then the regime in which a relativistic amplitude laser pulse $(a_0 > 1)$ irradiates an extended sub-critical plasma $(n_e < n_c)$. In order to illustrate the key features of the laser-plasma interaction in this regime, we have performed a two-dimensional (2D) particle-in-cell (PIC) simulation whose results are shown in Fig.~\ref{fig:2D_example}. In this simulation, a uniform sub-critical plasma with initial electron density $n_e = 0.01 n_c$ is irradiated by a laser beam with wavelength $\lambda = 1$ $\mu$m whose amplitude ramps up to $a_0 = 8.5$ and then remains constant. The laser pulse propagates along the $x$-axis and it is linearly polarized, with the laser electric field polarized in the $(x,y)$-plane. The ions were kept immobile in this simulation to distinguish more clearly the effect of the long laser pulse. Detailed parameters of the simulation are given in Appendix~\ref{Appendix_1}.

As the pulse enters the plasma, its ponderomotive force begins to expel some of the electrons out of the laser pulse in the transverse direction producing a channel. The un-neutralized ion charge generates a counteracting force that, in the example shown in Fig.~\ref{fig:2D_example}, prevents the channel from becoming fully evacuated. 

The laser pulse produces and maintains a steady-state channel if the pulse duration exceeds the characteristic electron response time. The snapshot of the electron density in Fig.~\ref{fig:2D_example} taken at 1 ps illustrates such a channel. The positively charged elongated channel generates quasi-static transverse and longitudinal electric fields shown in the lower-left panel of Fig.~\ref{fig:2D_example}. The fields have been averaged over ten laser periods and they are normalized to the electric field amplitude $E_0$ of a plane wave with $a_0 = 8.5$, which is essentially the amplitude of the electric field in the laser pulse. The time-averaged electric fields are relatively small compared to the amplitude of the oscillating laser electric field that is also present in the channel.

Once the steady-state channel structure is established, new electrons are continuously injected into the channel through the opening. This is particularly clear from the snapshot of the time-averaged current density, whose absolute value is shown in the upper-right panel of Fig.~\ref{fig:2D_example}. The injected electrons are accelerated and pushed forward by the laser pulse, producing a steady-state electron current in the channel. This current generates a quasi-static transverse magnetic field directed in and out of the plane of the simulation (along the $z$-axis). The return current flowing outside of the channel causes for the magnetic field to be localized inside the channel, as shown in the lower-right panel of Fig.~\ref{fig:2D_example}. The plotted time-averaged field is normalized to the magnetic field amplitude $B_0$ of a plane wave with $a_0 = 8.5$. The quasi-static magnetic field is also relatively weak in this regime, as it is less that 1\% of the magnetic field in the laser pulse.

\begin{figure}
	\centering
	\includegraphics[width=1.0\columnwidth]{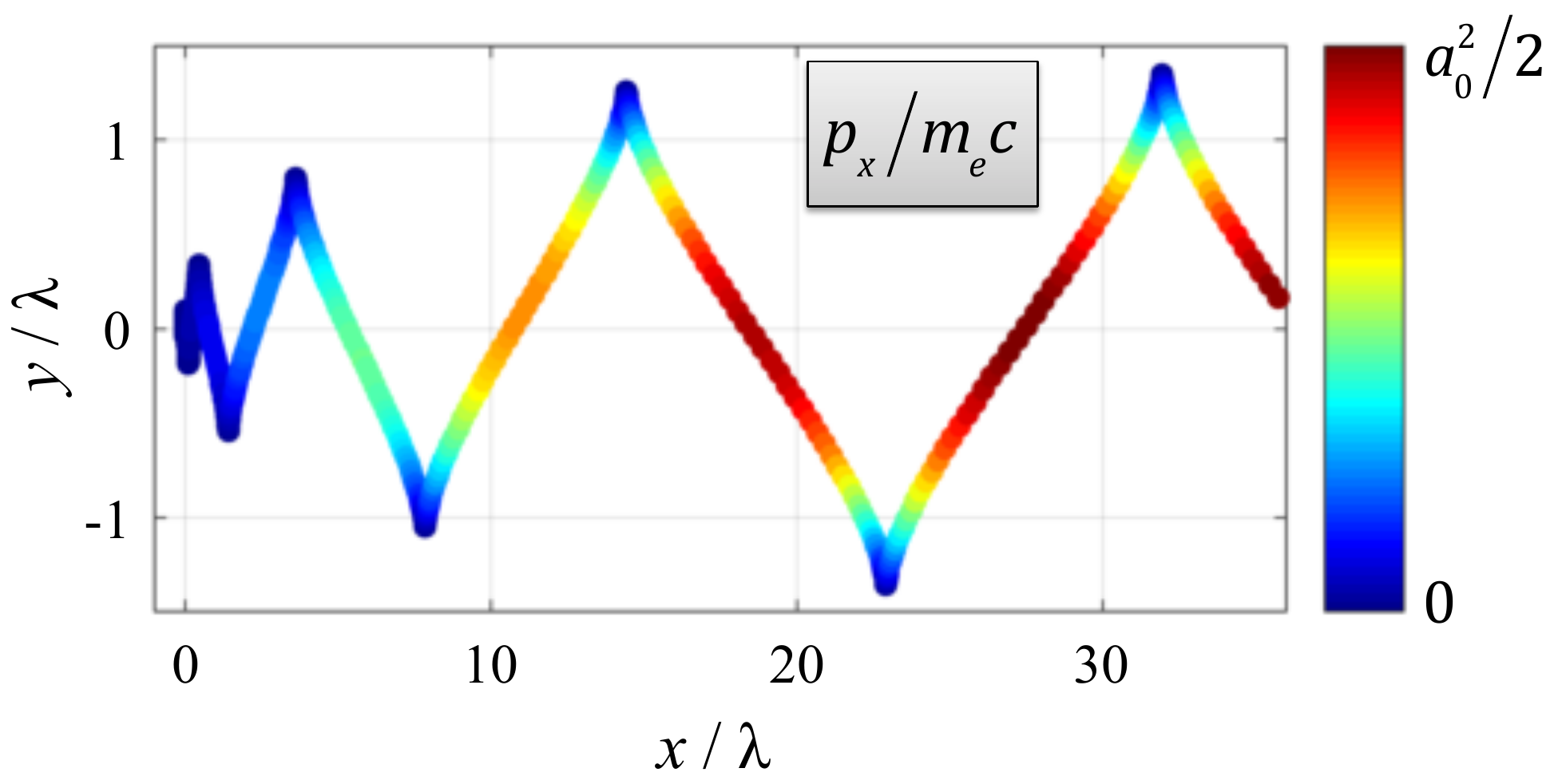} 
  \caption{Trajectory of an electron accelerated in a vacuum by a plane linearly polarized wave with $a_0 = 8.5$. Color-coded is the electron longitudinal momentum.} \label{Figure_DLA_vac_1}
\end{figure}

The presented example illustrates that, in a sub-critical plasma irradiated by a long laser pulse, electron acceleration takes place in a positively charged channel in the presence of extended transverse quasi-static electric and magnetic fields and a localized quasi-static longitudinal electric field. These fields are relatively small compared to the fields in the laser pulse, not exceeding a few percent. It is then somewhat unexpected that, for example, the transverse electric field can significantly enhance the electron energy gain~\cite{Pukhov1999}. As we show in the following sections, the transverse and longitudinal electric fields can synergistically enhance the electron energy gain, with a well-pronounced threshold, well beyond the energy gain that one would expect in the absence of these fields.


\section{Electron acceleration in a vacuum} \label{Sec-vac}

In order to provide the context for the discussion of the electron acceleration in the channel, we briefly review the key features of the electron motion in a vacuum where no static electric or magnetic fields are present. Specifically, we consider a single electron that is initially at rest. It is irradiated by a plane electromagnetic wave whose normalized amplitude gradually increases from zero to $a_0$. In what follows, we refer to this regime as the vacuum regime.

The electron moves according to the following equations:
\begin{eqnarray}
&& \frac{d {\bf{p}}}{d t} = - |e| {\bf{E}} - \frac{|e|}{\gamma m_e c} \left[ {\bf{p}} \times {\bf{B}} \right], \label{main_eq:1} \\
&& \frac{d {\bf{r}}}{d t} = \frac{c}{\gamma} \frac{{\bf{p}}}{m_e c}, \label{main_eq:2}
\end{eqnarray}
where $\gamma = \sqrt{1 + p^2 /m_e^2 c^2}$ is the relativistic factor, ${\bf{r}}$ and ${\bf{p}}$ are the electron position and momentum, and $t$ is the time. In the regime under consideration, ${\bf{E}}$ and ${\bf{B}}$ are the electric and magnetic fields of the wave. It is convenient to express these fields in terms of a normalized vector potential ${\bf{a}}$,
\begin{eqnarray}
&& {\bf{E}}_{wave} = - \frac{m_e c}{|e|} \frac{\partial {\bf{a}}}{\partial t}, \label{E_wave} \\
&& {\bf{B}}_{wave} = \frac{m_e c^2}{|e|} \left[ \nabla \times {\bf{a}} \right]. \label{B_wave}
\end{eqnarray}
In the case of a plane wave with wave-length $\lambda$ propagating along the $x$-axis, the vector potential is only a function of a normalized phase
\begin{equation} \label{xi}
\xi = \frac{2 \pi}{\lambda} \left(x - ct \right).
\end{equation}
Without any loss of generality, we assume that the laser electric field is polarized along the $y$-axis. In this case, the vector potential ${\bf{a}}$ only has a $y$-component. We consider a pulse with $a = a_0 F(\xi) \sin(\xi)$, where $a_0$ is the maximum normalized amplitude and $F(\xi)$ is a slowly varying envelope that ramps up from zero to unity.

Figure \ref{Figure_DLA_vac_1} shows an electron trajectory for a pulse with $a_0 = 8.5$ obtained by solving Eqs.~(\ref{main_eq:1}) and (\ref{main_eq:2}) numerically. It illustrates the qualitative change that takes place at relativistic wave amplitudes of $a_0 \geq 1$. At $a_0 \ll 1$, the dominant force experienced by the electron is the force from the electric field of the wave. As a result, the electron oscillates across the laser pulse, while  its longitudinal displacement is negligible. At $a_0 \geq 1$, the momentum oscillations induced by the laser electric field become relativistic and, as a consequence of this, the Lorentz force becomes important. This force causes longitudinal electron motion that leads to a trajectory shown in Fig.~\ref{Figure_DLA_vac_1}.

Even though the energy from the laser pulse is transferred to the transverse electron oscillations, the Lorentz force converts most of this energy into the longitudinal electron motion. Indeed, an analytical solution of Eqs.~(\ref{main_eq:1}) and (\ref{main_eq:2}) yields~\cite{Stupakov2001}
\begin{eqnarray}
&& p_x \left/ m_e c \right. = \left. a^2 \right/ 2, \label{vac_px}\\
&& p_y \left/ m_e c \right. = a. \label{vac_py}
\end{eqnarray}
The electron moves along a parabola in the momentum space, with $p_x \gg \left| p_y \right|$ for $a_0 \gg 1$. The maximum $\gamma$-factor that the electron can reach is thus given by
\begin{equation}
\gamma_{vac} = 1 + \left. a_0^2 \right/ 2.
\end{equation}
The change of the longitudinal electron momentum along the electron trajectory is shown in Fig.~\ref{Figure_DLA_vac_1}.

\begin{figure}
	\centering
	\includegraphics[width=1.0\columnwidth]{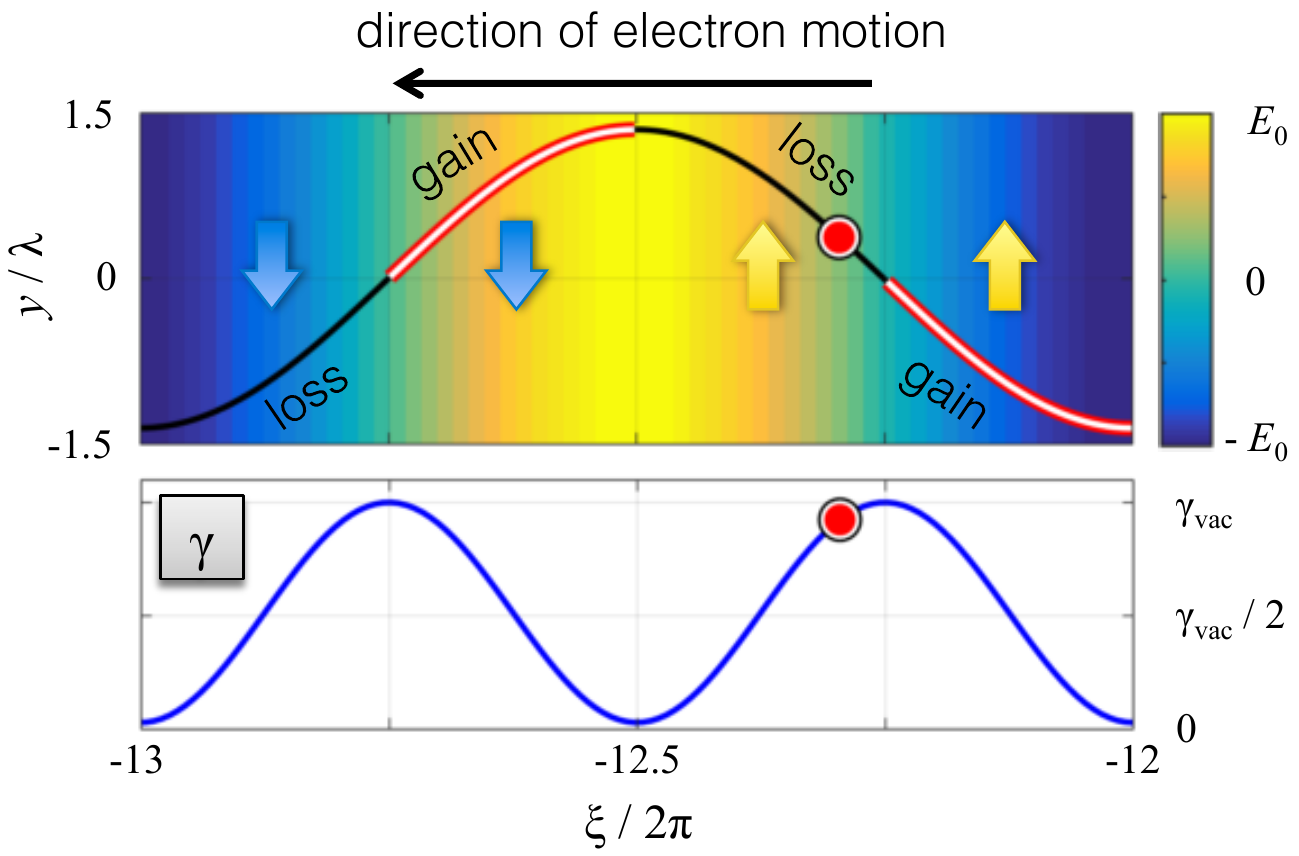} 
  \caption{Electron trajectory (upper panel) and the $\gamma$-factor (lower panel) as functions of the phase variable $\xi$ in a pulse with $a_0 = 8.5$. The color coding is the amplitude of the laser electric field. The arrows show the direction of the transverse electron velocity along the trajectory.} \label{Figure_DLA_vac_2}
\end{figure}

The energy gain is limited by the dephasing between the electron and the wave.  This aspect is illustrated in Fig.~\ref{Figure_DLA_vac_2}, where the electron trajectory and the $\gamma$-factor are shown as functions of the phase variable $\xi$. In the upper panel, the color-coding shows the amplitude of the laser electric field normalized to its maximum amplitude $E_0$. The phase of the wave at the electron location, $\xi$, continuously decreases, because the electron longitudinal velocity is less than the phase velocity $c$. The electron, shown with a circle, is slipping with respect to the laser wave fronts, thus moving to the left in the upper panel. At $\xi / 2 \pi = -12$, the electron transverse velocity (shown with arrows) is anti-parallel to the laser electric field, so the electron is gaining energy from the laser pulse. As the electron continues to slip, the laser electric field eventually changes its sign at $\xi / 2 \pi = -12.25$ and becomes positive. At this point, the electron transverse velocity is still positive and the electron starts to lose its energy. This positive field continues to reduce the transverse electron velocity until it becomes negative at  $\xi / 2 \pi = -12.5$, at which point the electron begins to gain energy again. The electron experiences energy gain twice every laser cycle, as evident from the plot of the $\gamma$-factor in Fig.~\ref{Figure_DLA_vac_2}. It is worth pointing out that the electron comes momentarily to a complete stop also twice every laser cycle. The longitudinal distance that the electron travels between the stops to achieve $\gamma = \gamma_{vac}$ is much longer than the laser wavelength due to the fact that the electron is moving forward and it can be estimated as $\gamma_{vac} \lambda /4$. 

The discussed mechanism of electron acceleration and energy gain in the vacuum regime suggests that, in principle, there are two alternatives for further enhancing the electron energy gain. One option is to decrease the dephasing between the electron and the wave, allowing the transverse electron velocity to remain antiparallel to the laser electric field for longer in terms of the actual time. The second option is to change the oscillations of the transverse velocity, so that there is a net energy gain by the electron after a laser cycle. In what follows, we show how these scenarios can be realized utilizing relatively weak static longitudinal and transverse electric fields. 


\section{Role of the quasi-static longitudinal electric field}

As discussed in Sec.~\ref{Sec-channel}, one of the key features of the steady-state channel is the continuous injection of new electrons through the channel opening with a quasi-static longitudinal electric field. This aspect raises the question of the impact of the quasi-static longitudinal electric field on subsequent electron energy gain during acceleration by the laser pulse along the channel.

\begin{figure}
	\centering
	\includegraphics[width=0.8\columnwidth]{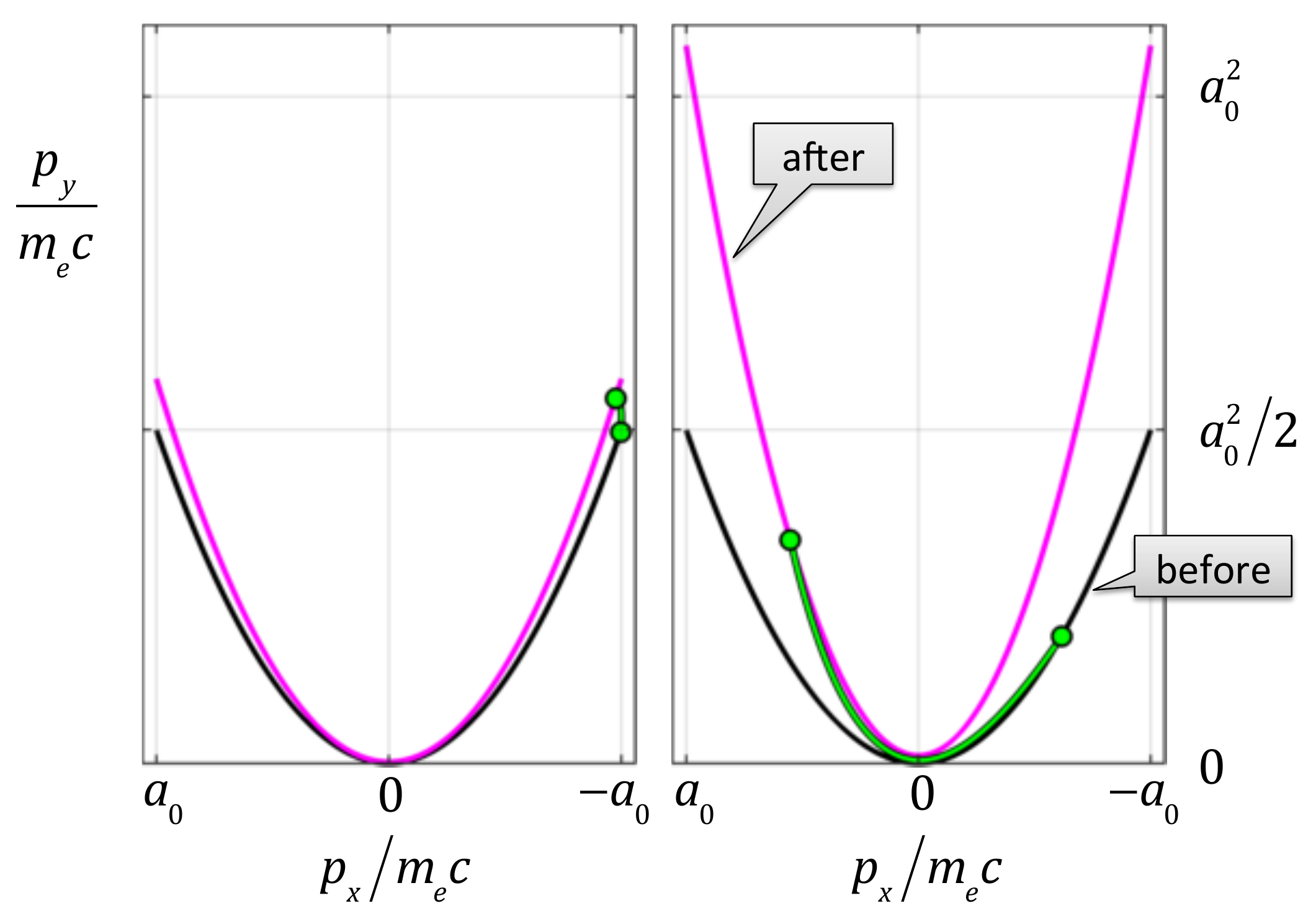} 
  \caption{Momentum of an electron before, during, and after crossing a region with a static electric field $E_* = -0.05 E_0$ that is $\Delta x = 2 \lambda$ long. The only difference between the two cases is the location of the field region.} \label{Figure_DLA_Ez_1}
\end{figure}

In order to pinpoint the effect of the longitudinal electric field, we begin by considering electron motion in a vacuum using the same setup as in Sec.~\ref{Sec-vac}, but with an added narrow region $\Delta x$ of a uniform longitudinal static electric field $E_*$. Figure \ref{Figure_DLA_Ez_1} shows numerical solutions of Eqs.~(\ref{main_eq:1}) and (\ref{main_eq:2}) for $E_* = -0.05 E_0$ and $\Delta x = 2 \lambda$, where $E_0$ is the electric field amplitude of the laser pulse. The laser pulse amplitude is again $a_0 = 8.5$ following a gradual initial ramp-up. The difference between the two examples shown in Fig.~\ref{Figure_DLA_Ez_1} is the location of the field region along the electron trajectory. In the first example (left panel), the interaction takes place when the electron momentum is close to its maximum value. The subsequent change in the maximum electron momentum after the interaction is insignificant. This result is in agreement with the expectation that the localized field we are considering is too weak to transfer significant energy to the electron. In the second example (right panel), the interaction takes place as the electron comes to a stop. In stark contrast to the previous example, the maximum electron energy more than doubles during the motion in the wave after the interaction.

The key role of the longitudinal electric field in these examples is not to directly increase the electron energy, but rather to decrease the electron dephasing with the wave. We define the dephasing,
\begin{equation} \label{R}
R \equiv - \frac{1}{\omega} \frac{d \xi}{d \tau},
\end{equation}
as the rate at which the wave phase at the electron location, $\xi$, changes with proper time, $\tau$, defined as 
\begin{equation} \label{tau_def}
\frac{d \tau}{dt}  = \frac{1}{\gamma}.
\end{equation}
It follows directly from the definition of $\xi$ given by Eq.~(\ref{xi}) that
\begin{equation} \label{R2}
R = \gamma - p_x \left/ m_e c \right. .
\end{equation}
The quantity $\gamma - p_x / m_e c$ is conserved by Eqs.~(\ref{main_eq:1}) and (\ref{main_eq:2}) in the absence of the static electric field, which means that the dephasing defined by Eq.~(\ref{R}) is constant before and after the interaction with the longitudinal field. It can also be shown that during the interaction the static field reduces the dephasing at the rate~\cite{Robinson2013,Arefiev2015}
\begin{equation}
\frac{d R}{d \xi}= \frac{|e| E_*}{m_e c}
\end{equation}
that is independent of the laser pulse amplitude.  


The reduced dephasing allows the electron to spend more time being accelerated by the wave and thus gain more energy from the laser pulse. As shown in Fig.~\ref{Figure_DLA_vac_2} of Sec.~\ref{Sec-vac}, the acceleration continues before the phase decreases by $\Delta \xi  = -\pi/2$. A reduction in the dephasing then means that the corresponding proper time interval $\Delta \tau$ of electron acceleration is indeed longer [see Eq.~(\ref{R})]. The maximum $\gamma$-factor that the electron with reduced dephasing achieves~\cite{Robinson2013,Arefiev2015} is
\begin{equation} \label{gamma_long}
\gamma_{\max} \approx R^{-1} \gamma_{vac}.
\end{equation}

\begin{figure}
	\centering
	\includegraphics[width=0.9\columnwidth]{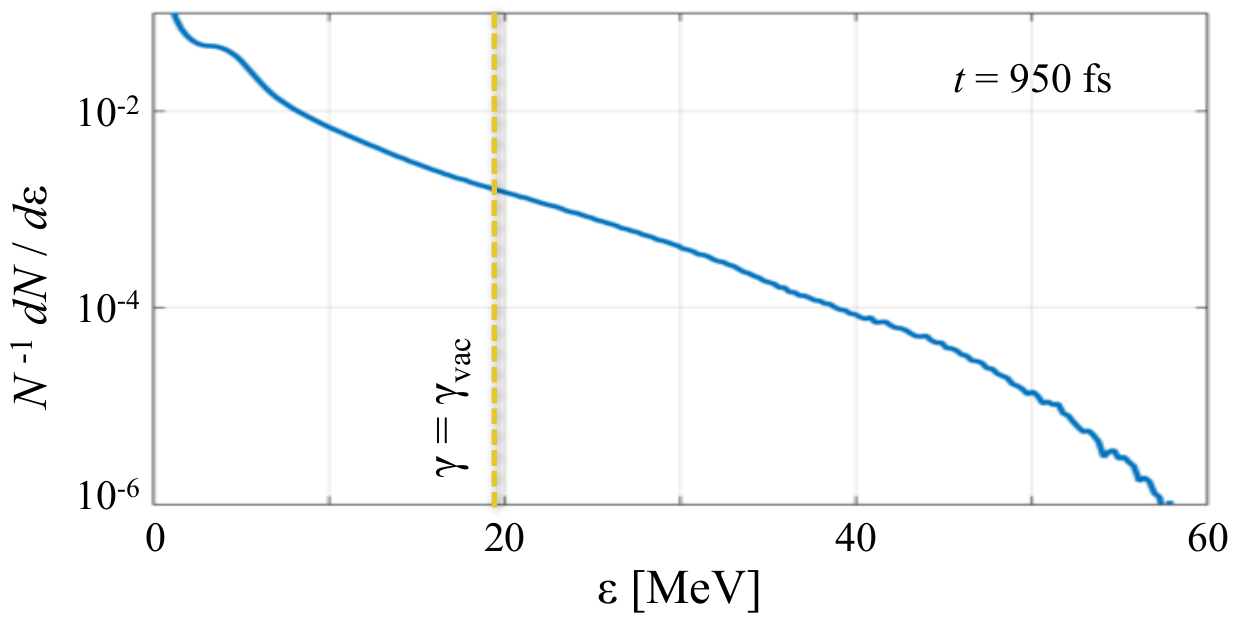} 
  \caption{Snapshot of a normalized electron spectrum at $t = 950$ fs from a 2D PIC simulation with $a_0 = 8.5$ and $n_e = 0.05 n_c$.} \label{Figure_Spectrum}
\end{figure}

Not all parts of the electron trajectory are equivalent in terms of the dephasing change induced by the longitudinal field. The reason for that is the strong dependence of the dephasing on the longitudinal momentum. For simplicity, let us assume that that the electric field instantaneously increases the longitudinal momentum by $\Delta p_x$. Prior to the interaction, we have
\begin{equation}
R = \gamma - p_x \left/ m_e c \right. \approx \left. p_y^2 \right/ 2 p_x = 1.
\end{equation}
Here we used expressions (\ref{vac_px}) and (\ref{vac_py}) for the electron momentum in the vacuum regime. Assuming that $\Delta p_x$ is smaller than $p_x$ and $p_y$, we find from Eq.~(\ref{R2}) that the dephasing is reduced to
\begin{equation}
R \approx 1 - \left. \Delta p_x \right/ p_x.
\end{equation}
Evidently, the most favorable parts of the electron trajectory for reducing the dephasing are those parts where the longitudinal momentum is the lowest. This aspect is at work in the two examples shown in Fig.~\ref{Figure_DLA_Ez_1}, where the dephasing drops to $R = 0.47$ on the right and only to $R = 0.87$ on the left. The resulting energy increase in both cases is in good agreement with Eq.~(\ref{gamma_long}).

\begin{figure}
	\centering
	\includegraphics[width=1.0\columnwidth]{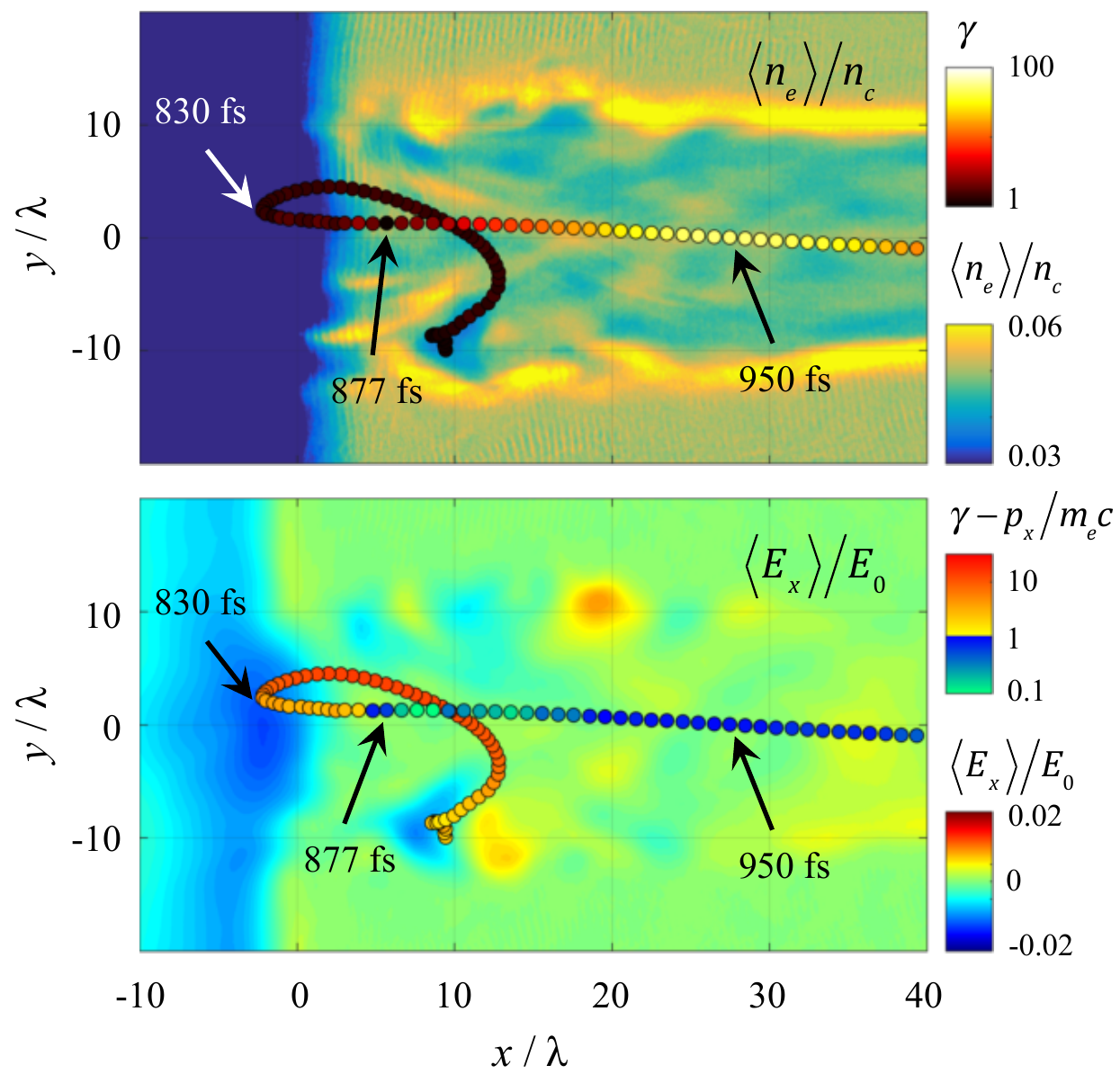} 
  \caption{Trajectory of an accelerated electron in a channel. Upper panel shows the trajectory, with color-coded $\gamma$, on top of the time-averaged electron density profile. The lower panel shows the trajectory, with the color-coded dephasing rate $R = \gamma - p_x/m_e c$, on top of the time-averaged longitudinal electric field. The field and the density are averaged over ten laser periods at 850 fs.} \label{Figure_DLA_Ez_2}
\end{figure}

In order to examine how this mechanism is realized in a plasma channel, we have performed a 2D PIC simulation in which a long laser pulse with an electric field polarized perpendicular to the plane of the simulation irradiates a plasma slab with $n_e = 0.05 n_c$. Detailed parameters of the simulation are given in Appendix~\ref{Appendix_1}. The normalized laser amplitude is $a_0 = 8.5$, so the $\gamma$-factor in the vacuum regime would be limited by $\gamma_{vac} \approx 37$. However, the electron spectrum in the plasma, as seen in Fig.~\ref{Figure_Spectrum}, has a pronounced energetic electron tail with $\gamma > \gamma_{vac}$. 

A trajectory of one of the energetic electrons is shown in Fig.~\ref{Figure_DLA_Ez_2} together with a time-averaged electron density profile and longitudinal electric field. The electron is injected into the channel near the opening and initially moves against the laser pulse. The electron typical $\gamma$-factor at this stage is comparable to $a_0$. The quasi-static negative electric field, that is present in the channel opening, gradually turns the electron around. As the electron starts to move in the direction of the laser pulse propagation, the regime becomes similar to that considered earlier in this Section. The longitudinal electric field causes the dephasing rate to decrease, as shown in the lower panel of Fig.~\ref{Figure_DLA_Ez_2}. Similarly to the simplified example, the significant drop in the dephasing occurs as the $\gamma$-factor reaches its minimum at $t = 877$ fs. 

\begin{figure}
	\centering
	\includegraphics[width=0.9\columnwidth]{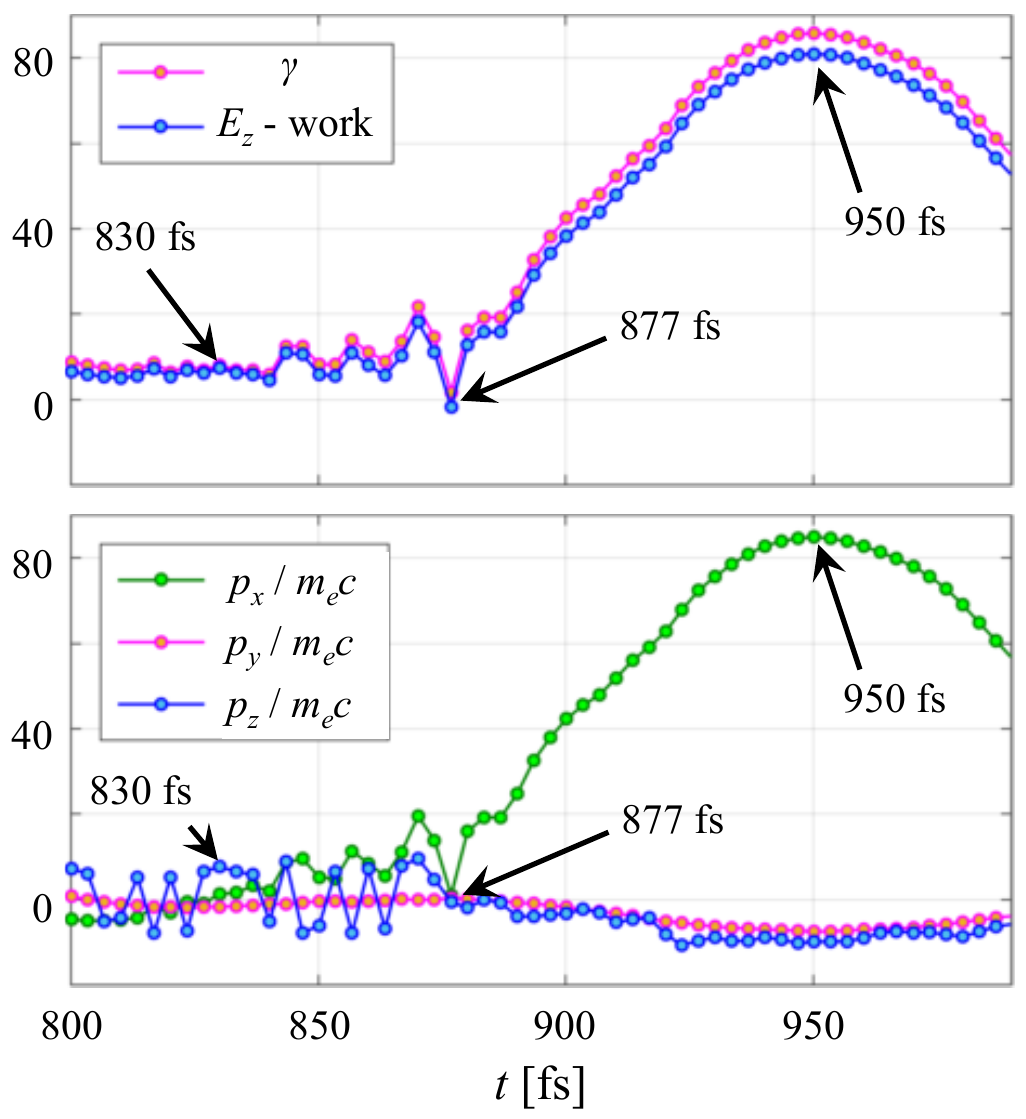} 
  \caption{Electron $\gamma$-factor and momentum components as functions of time for the electron trajectory shown in Fig.~\ref{Figure_DLA_Ez_2}. The upper panel also shows the integrated contribution to the $\gamma$-factor from the work done by the laser electric field.} \label{Figure_DLA_Ez_3}
\end{figure}

The energy increase takes place as the electron continues its longitudinal motion into the channel after leaving the region with a negative quasi-static electric field. The reduced dephasing is evident not only from the color-coding in the lower panel of Fig.~\ref{Figure_DLA_Ez_2}, but also from the lack of oscillations in $p_z$ in Fig.~\ref{Figure_DLA_Ez_3}. The electron $\gamma$-factor peaks at $t = 950$ fs, roughly 20 $\mu$m after the energy increase began. It exceeds $\gamma_{vac}$ by a factor of 2.3. Most of the energy comes directly from the laser electric field and not from the longitudinal field, as shown in the upper panel of Fig.~\ref{Figure_DLA_Ez_3}. The laser contribution to the $\gamma$-factor is calculated by integrating the work done by $E_z$ and normalizing it to $m_e c^2$.

\subsection*{Role of the superluminal phase velocity}

We have so far assumed that the phase velocity in the laser pulse, $v_{ph}$, is equal to the speed of light, $c$. However, the plasma in the channel and the channel itself cause dispersion, making the phase velocity superluminal, $v_{ph} > c$. The effect of the superluminosity is faster dephasing between the electron and the wave and, as a result, lower energy gain from the wave. 

The superluminosity must be taken into account if it leads to a significant change in the dephasing. In general, the dephasing is determined by by the difference $v_{ph} - v_x$. We can then conclude that the superluminosity becomes important for $v_{ph} - c \geq c - v_x$, whereas we can set $v_{ph} = c$ for $v_{ph} - c \ll c - v_x$. Using the definitions $R = \gamma - p_x / m_e c$ and $v_x = p_x / \gamma m_e$, we find that
\begin{equation} \label{v_ph_1_2}
c - v_x = \left. c R \right/ \gamma.
\end{equation}
In the case when $v_{ph} = c$, the $\gamma$-factor is given by Eq.~(\ref{gamma_long}). Taking into account this estimate and the expression for $\gamma_{vac}$, we find that the role of the superluminosity is negligible for~\cite{Robinson2015}
\begin{equation} \label{v_ph_cond_2}
\frac{v_{ph} - c}{c} \ll \frac{R^2}{a_0^2}.
\end{equation}

Ultimately, the superluminosity sets a limit on the energy gain from the wave. It follows from Eq.~(\ref{v_ph_cond_2}) that there is a critical value for $R = \gamma - p_x / m_e c$, given by
\begin{equation} \label{critical_dephasing}
R_* = a_0 \sqrt{\frac{v_{ph} - c}{c}}.
\end{equation}
For $R \ll R_*$, the dephasing between the electron and the wave is determined primarily by the difference $v_{ph} - c$ and the difference between the electron longitudinal velocity and the speed of light is unimportant. This means that even if the longitudinal electric field lowers $R$ well below $R_*$, the maximum $\gamma$-factor that the electron can achieve will be approximately limited by
\begin{equation} \label{v_ph_cond_3}
\gamma_{max} \approx \sqrt{\frac{1}{a_0} \frac{c}{v_{ph} - c}} \gamma_{vac},
\end{equation}
which was estimated from Eq.~(\ref{gamma_long}) by setting $R = R_*$.

\begin{figure}
	\centering
	\includegraphics[width=1.0\columnwidth]{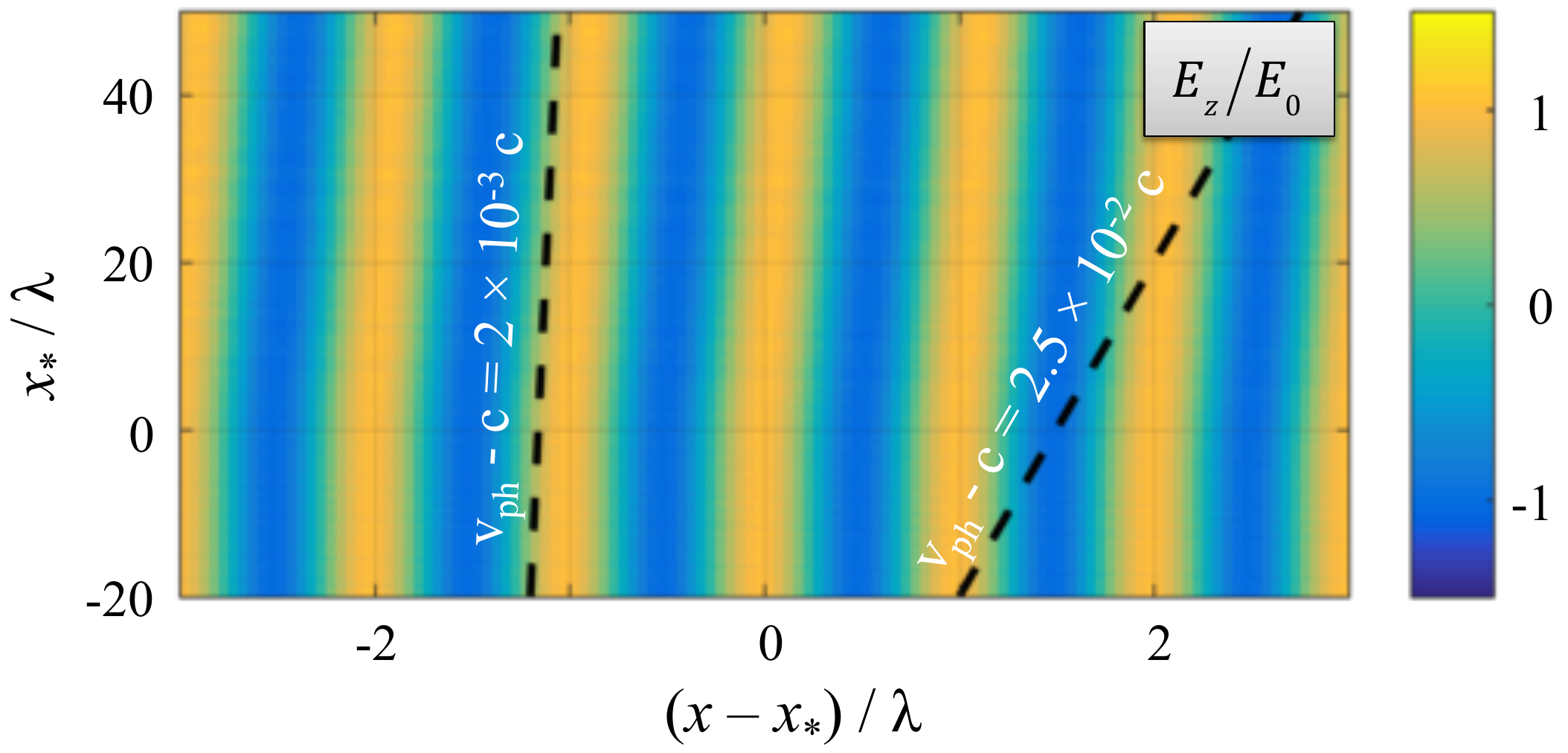} 
  \caption{Wavefronts in a window moving with the speed of light along the axis of the channel. Here $x_*$ is the location of the center of the window. The field is normalized to the electric field amplitude $E_0$ of a plane wave with $a_0 = 8.5$. The dashed lines represent wavefronts moving with two different superluminal phase velocities $v_{ph}$. } \label{Figure_v_ph}
\end{figure}

In order to examine the conditions (\ref{v_ph_cond_2}) and (\ref{v_ph_cond_3}) for the simulation shown in Fig.~\ref{Figure_DLA_Ez_2}, we have plotted the wavefronts in a window moving with the speed of light along the axis of the channel. Figure~\ref{Figure_v_ph} shows the wave electric field $E_z$ on the axis of the channel in a window that is $6 \lambda$ wide. The center of the window, located at $x = x_*$, slides the entire length of the channel shown in Fig.~\ref{Figure_DLA_Ez_2} (note that $\lambda = 1$ $\mu$m in the simulation). The two dashed lines indicate wavefront slopes for $v_{ph} - c = 2 \times 10^{-3} c$ and $v_{ph} - c = 2.5 \times 10^{-2} c$. Inside the channel we then approximately have $v_{ph} - c = 2 \times 10^{-3} c$.  The corresponding critical dephasing is $R_* \approx 0.38$ and the maximum $\gamma$ that can be attained by accelerating electrons is $\gamma_{\max} \approx 2.6 \gamma_{vac}$, which translates into $\gamma_{\max} \approx 98$ for $a_0 = 8.5$. 

In the light of these estimates, a closer look at the electron data shown in Figs.~\ref{Figure_DLA_Ez_2} and \ref{Figure_DLA_Ez_3} reveals that the superluminosity likely plays a role in limiting the electron energy gain. The dephasing rate $R$ shown in Fig.~\ref{Figure_DLA_Ez_2} falls to $R \approx 0.1$ along the electron trajectory, which could lead to a factor of ten increase of the $\gamma$-factor compared to $\gamma_{vac}$ for $v_{ph} = c$. However, our estimates show that reduction of the dephasing below $R_* \approx 0.38$ no longer leads to an enhancement of the $\gamma$-factor, capped at $\gamma_{\max} \approx 98$, because of the superluminosity. This is in good agreement with the $\gamma$-profile shown in the upper panel of Fig.~\ref{Figure_DLA_Ez_3}.

Finally, it is worth pointing out that heating of the bulk electrons in the channel to relativistic energies significantly reduces the superluminosity. The phase velocity of a linear electromagnetic wave propagating through a cold non-relativistic plasma with electron density $n_e$ is $v_{ph} / c = \left( 1 - n_e/n_c \right)^{-1/2}$. We then should expect $(v_{ph} - c) / c \approx 2.6 \times 10^{-2}$ for $n_e = 0.05 n_c$ used in the simulation. The corresponding wavefront should follow the right dashed line in Fig.~\ref{Figure_v_ph}. The actual $v_{ph} - c$ is smaller by an order of magnitude, which clearly indicates that the critical density that determines the phase velocity is effectively lowered by an order of magnitude due to relativistic motion induced in the channel by the laser pulse. This so-called relativistic transparency enables the electron energy gain enhancement by reducing the superluminosity and thus reducing the critical dephasing given by Eq.~(\ref{critical_dephasing}).


\section{Role of the quasi-static transverse electric field}

\begin{figure}
	\centering
	\includegraphics[width=0.9\columnwidth]{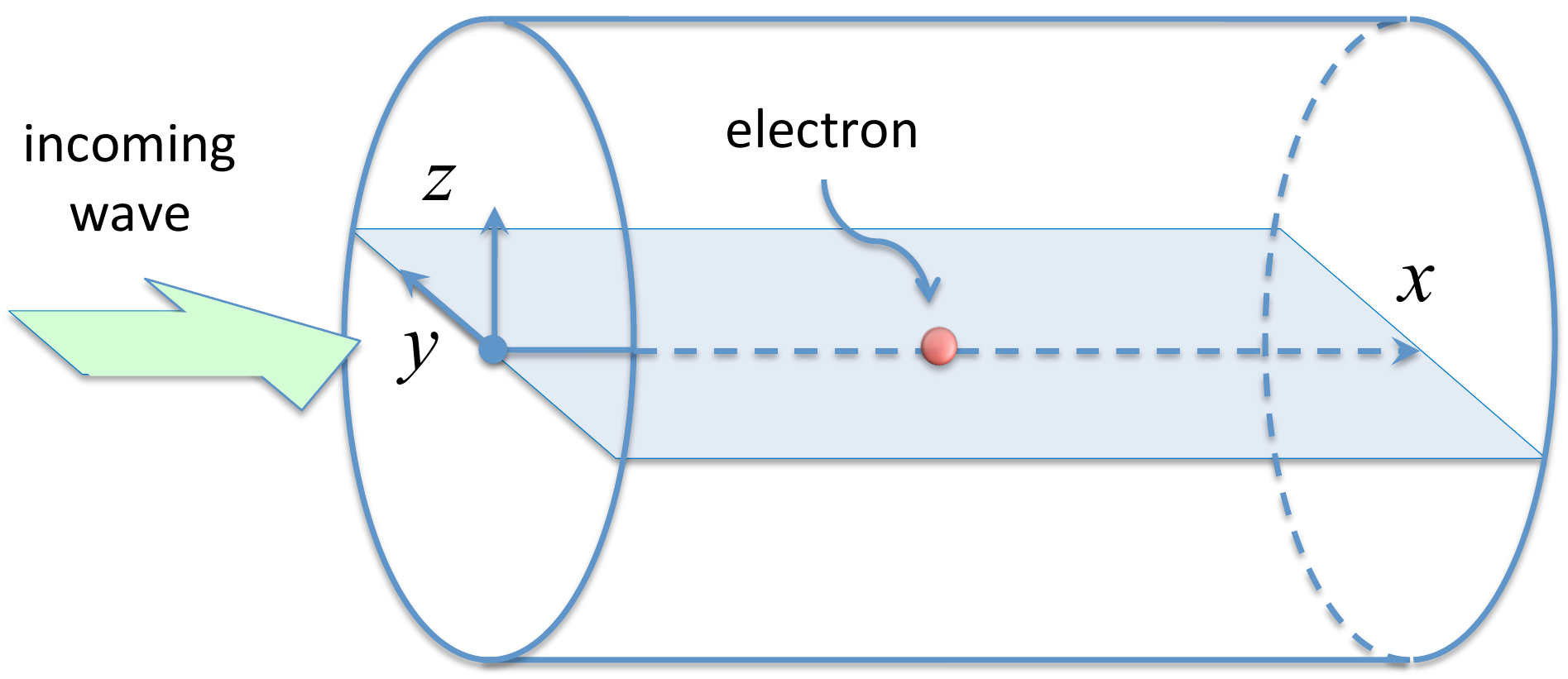} 
  \caption{Setup  used to examine the role of the transverse static electric field. The laser electric field is polarized in the $(x,y)$-plane.} \label{Figure_cyl_setup}
\end{figure}

Inside the channel, the electron acceleration by the laser pulse takes place in the presence of a relatively weak quasi-static transverse electric field. In order to examine the role played by such a field, we consider just a single electron irradiated by a plane electromagnetic wave in a fully evacuated cylindrical ion channel shown in Fig.~\ref{Figure_cyl_setup}.

The electron dynamics in this model is described by Eqs.~(\ref{main_eq:1}) and (\ref{main_eq:2}), where the electric and magnetic fields are given. These fields are a superposition of the fields of the wave given by Eqs.~(\ref{E_wave}) and (\ref{B_wave}) and the electric field of the channel, 
\begin{eqnarray}
&& E_y^{chan} = \left. m_e \omega_{p0}^2 y \right/ 2 |e|, \\
&& E_z^{chan} = \left. m_e \omega_{p0}^2 z \right/ 2 |e|, 
\end{eqnarray}
where $\omega_{p0} \equiv \sqrt{4 \pi n_0 e^2 / m_e}$. We have assumed for simplicity that the channel is an infinitely long uniform cylinder that consists of immobile singly-charged ions with density $n_0$. 

It can be shown that, as the electron moves along the channel, the following quantity remains conserved
\begin{eqnarray}
\mathcal{I}_0 = \gamma - \frac{p_x}{m_e c} + \frac{\omega_{p0}^2}{4 c^2} \left( y^2 + z^2 \right) = \mbox{const}. \label{Q_def}
\end{eqnarray}
This relation indicates that amplification of electron oscillations across the channel leads to a reduction of $\gamma - p_x / m_e c$. Since $p_x$ is the dominant component of the electron momentum, a significant reduction of $\gamma - p_x / m_e c$ implies a significant increase of $p_x$. Therefore, the relation (\ref{Q_def}) formally points at a direct connection between enhancement of transverse oscillations and a significant increase of the longitudinal electron momentum.

The integral of motion (\ref{Q_def}) sets an upper limit for the amplitude of the transverse oscillations across the channel. The value of $\mathcal{I}_0$ is determined by the initial conditions, so that $\mathcal{I}_0 = 1$ for an on-axis electron that is initially at rest. We find from Eq.~(\ref{Q_def}) that the amplitude of the electron oscillations cannot exceed 
\begin{eqnarray}
r_{\max} = \frac{\lambda}{\pi} \frac{\omega}{\omega_{p0}} \label{Q_max_2}
\end{eqnarray}
for $\mathcal{I}_0 = 1$. The amplitude of the oscillations approaches $r_{\max}$ as $p_x/m_e c \rightarrow \infty$ and $\gamma - p_x/m_e c \rightarrow 0$. 

It is important to distinguish transverse electron oscillations in the $y$ and $z$ directions, because the motion in the $y$-direction is driven by the laser electric field, whereas the motion in the $z$-direction is only affected by the electric field of the channel. We first focus on the key aspects of the driven electron oscillations. We consider an electron that has no initial displacement or momentum in the $z$-direction, so that the electron trajectory driven by the laser pulse remains flat, with the electron moving only in the $(x,y)$-plane.

\subsection*{Driven motion across the channel}

The equation for the electron motion across the channel can be written in the form that resembles an equation for a driven oscillator~\cite{Arefiev2014}:
\begin{eqnarray} \label{perp_osc}
\frac{d^2 y}{d \tau^2} + \Omega^2 y = c \frac{d a}{d \tau},
\end{eqnarray}
where
\begin{equation}
\Omega \equiv \sqrt{ \gamma / 2 } \omega_{p0}
\end{equation}
and $\tau$ is the proper time defined by Eq.~(\ref{tau_def}). The wave amplitude is a function of the phase variable $\xi$ that changes at the rate
\begin{equation}
- \frac{1}{\omega} \frac{d \xi}{d \tau} = \gamma - \frac{p_x}{m_e c}. 
\end{equation}

There are two characteristic frequencies in this case: the frequency of the oscillations induced by the field of the channel electric field and the frequency of the oscillations induced by the field of the laser pulse. The former is the natural frequency $\Omega$. The latter is equal to $\omega$ in the vacuum regime, because $\gamma - p_x / m_e c = 1$ and, as a result, $d \xi / d \tau = - \omega$. According to Eq.~(\ref{Q_def}), the relation $d \xi / d \tau \approx - \omega$ holds in the presence of the transverse electric field as well while the amplitude of the transverse oscillations is much smaller than $r_{max}$.

At low channel densities, such that $\Omega \ll \omega$, the channel electric field has very little impact on the electron motion. The frequency mismatch means that a resonant interaction that can lead to an amplification of the transverse oscillations is not possible. The electron motion is essentially the same as in the vacuum regime, with $\gamma \approx \gamma_{vac}$. The characteristic amplitude of the transverse oscillations, estimated as  $\Delta y \approx c a_0 / \omega$ from Eq.~(\ref{perp_osc}), is indeed much smaller than $r_{\max}$ for $\Omega \ll \omega$ and $\gamma \approx \gamma_{vac}$.

\begin{figure}
	\centering
	\includegraphics[width=1.0\columnwidth]{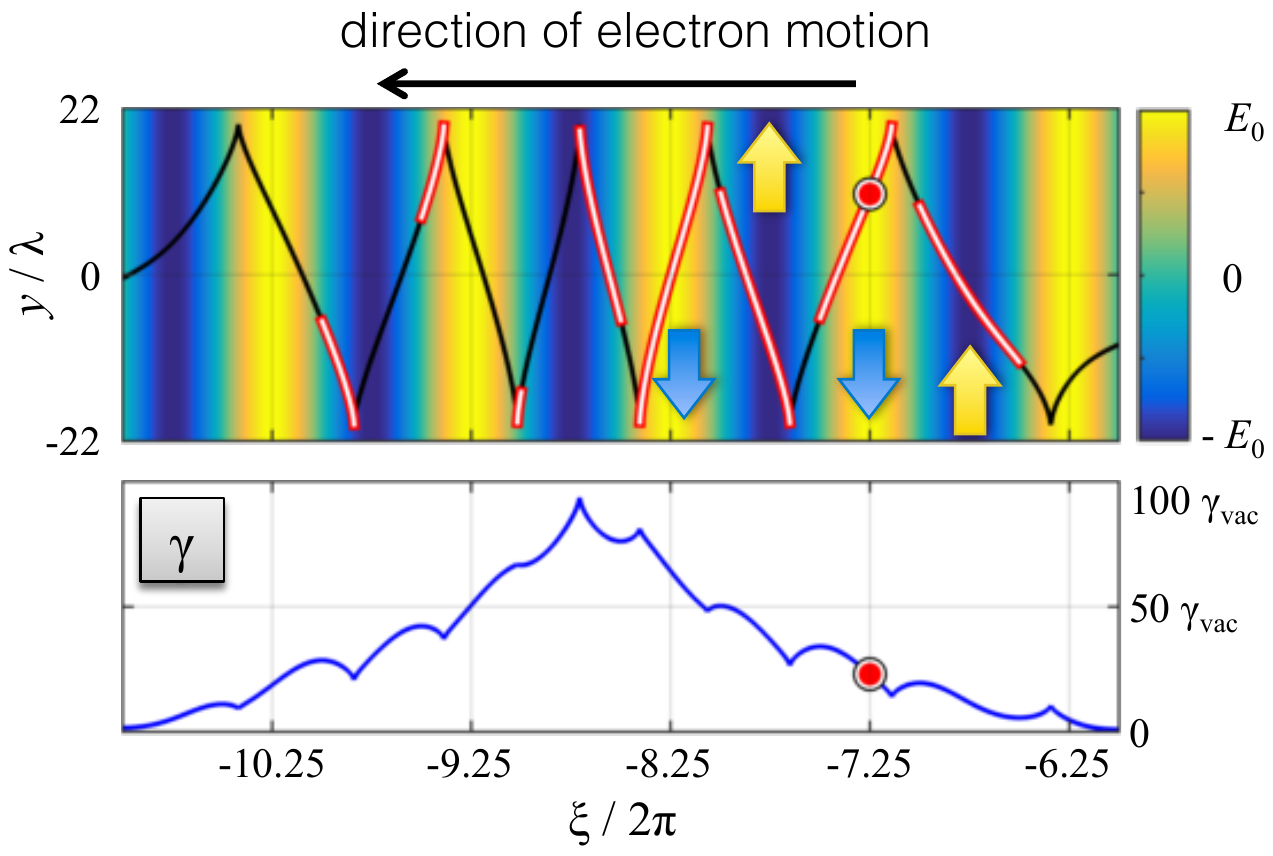} 
  \caption{Trajectory (upper panel) and the $\gamma$-factor (lower panel) of an electron irradiated by a plane wave with $a_0 = 11.5$ in a channel whose ion density corresponds to $\omega_{p0}/\omega = 0.016$. The color coding is the amplitude of the laser electric field. The arrows show the direction of the transverse electron velocity along the trajectory.} \label{Figure_DLA_vac_3}
\end{figure}

At higher channel densities for which the two characteristic frequencies become comparable, $\Omega \sim \omega$, the channel electric field becomes capable of significantly changing the phase of the transverse oscillations, which causes their amplitude to grow. The change of phase after significant amplification of the transverse oscillations has already taken place is illustrated in Fig.~\ref{Figure_DLA_vac_3}, where the electron trajectory and the $\gamma$-factor are shown as functions of the phase variable $\xi$. In this example, the electron is initially at rest on the axis of the channel ($x = y = z = 0$) whose ion density corresponds to $\omega_{p0}/\omega = 0.016$. The electron is irradiated by a plane wave with $a(\xi) = a_0 \cos(\xi)$ and $a_0 = 11.5$. The upper panel of Fig.~\ref{Figure_DLA_vac_3} shows that the channel electric field changes the oscillations of the transverse electron velocity, allowing it to remain anti-parallel to the laser electric field over extended segments of the electron trajectory marked with white. As a result, there is a net energy gain each laser cycle for $-6.25 > \xi/2 \pi > -8.75$ that leads to a significant increase of the relativistic $\gamma$-factor well above what can be achieved in the vacuum regime illustrated in Fig.~\ref{Figure_DLA_vac_2}. Note that $(r_{\max} - \max |y|) / r_{\max} \approx 5 \times 10^{-5}$ along this segment of the electron trajectory. The example in Fig.~\ref{Figure_DLA_vac_3} confirms that a significant increase of the $\gamma$-factor goes hand in hand with the enhancement of the transverse oscillations.

\begin{figure}
	\centering
    	\subfigure{\includegraphics[width=0.9\columnwidth]{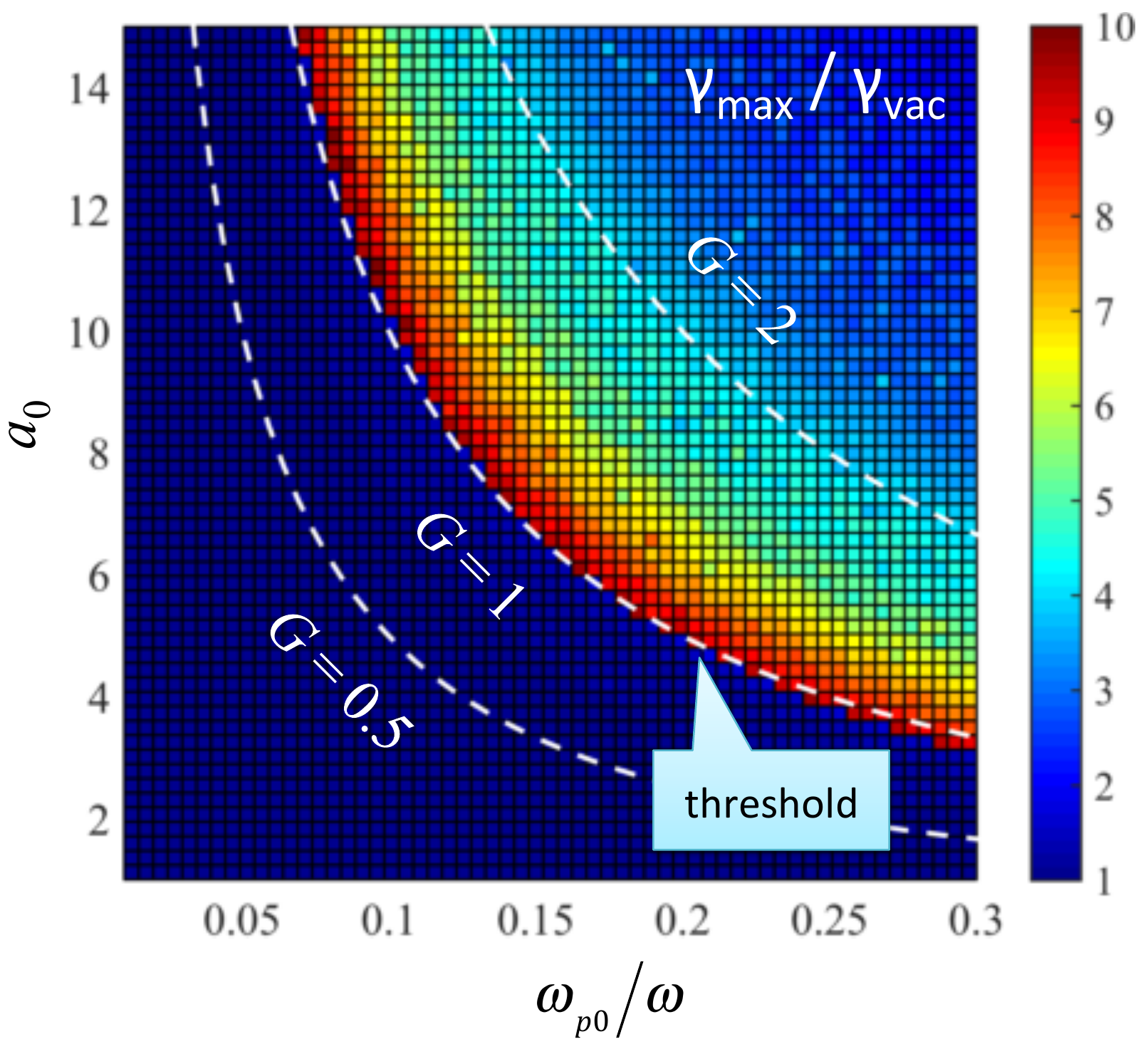} } \\
	\subfigure{\includegraphics[width=0.8\columnwidth]{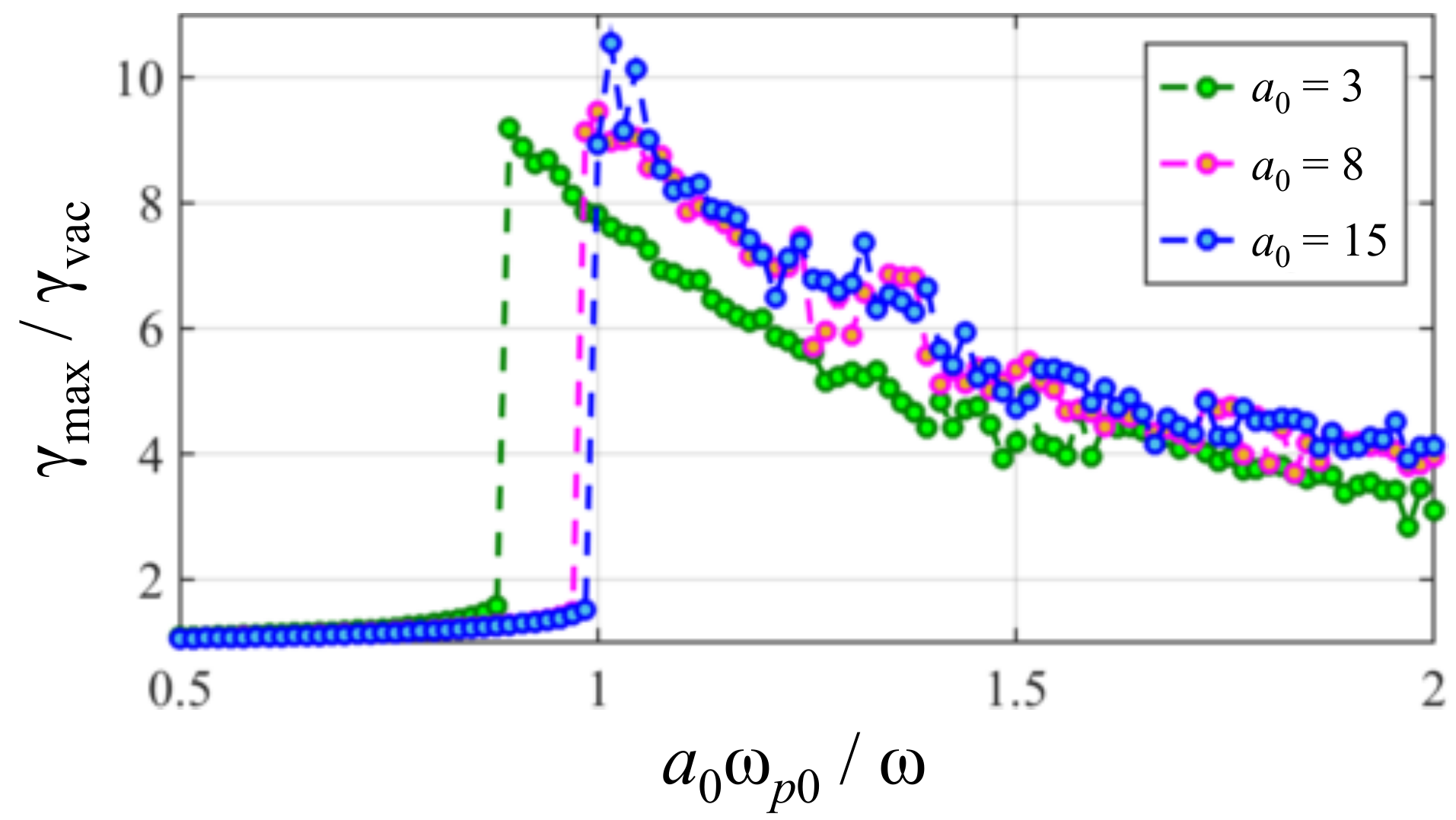}} 
  \caption{Maximum $\gamma$-factor attained by an electron irradiated by a plane wave with amplitude $a_0$ in a channel with ion density $n_0$ ($\omega_{p0} = \sqrt{4 \pi n_0 e^2 / m_e}$). In the upper panel, $G = a_0 \omega_{p0} / \omega$ and $\gamma_{vac} = 1 + a_0^2/2$.} \label{Invited_PoP_Figure_8}
\end{figure}

In order to determine the conditions for the energy enhancement, we have performed a parameter scan solving Eqs.~(\ref{main_eq:1}) and (\ref{main_eq:2}) numerically for a wide range of parameters $a_0$ and $\omega_{p0}/\omega$ that specify the fields acting on the electron. In all the cases, the electron is initially at rest on the axis of the channel and the laser pulse amplitude gradually ramps up as 
\begin{equation}
a(\xi) =
   \begin{cases}
      a_0 \exp[-(\xi-\xi_0)^2/2 \sigma^2] \sin(\xi),  & \text{for $\xi >  \xi_0$;}\\
      a_0 \sin(\xi),  & \text{for $\xi \leq  \xi_0$}
   \end{cases} \label{a}
\end{equation}
where $\xi_0 = -50$ and $\sigma = 20$. The maximum $\gamma$-factor, $\gamma_{\max}$, that the electron attains for different sets of parameters $a_0$ and $\omega_{p0}/\omega$ is shown in Fig.~\ref{Invited_PoP_Figure_8}, where $\gamma_{\max}$ is normalized to the maximum $\gamma$-factor in the vacuum regime, $\gamma_{vac}$, for the same laser amplitude $a_0$. 

The energy enhancement has a sharp threshold, as further illustrated by the lineouts in Fig.~\ref{Invited_PoP_Figure_8} for three different values of $a_0$. The lineouts for $a_0 = 8$ and $a_0 = 15$ indicate that the threshold and the energy gain above the threshold are determined by a single dimensionless combination 
\begin{equation} \label{G}
G \equiv a_0 \omega_{p0} / \omega.
\end{equation}
The contours of constant $G$ plotted in the upper panel of Fig.~\ref{Invited_PoP_Figure_8} with dashed lines confirm that this is indeed a general trend for $a_0 > 5$. 

The existence of the threshold that depends only on $a_0 \omega_{p0} / \omega$ was first discovered in Ref.~[\onlinecite{Arefiev2012}] and then explored in more detail in Ref.~[\onlinecite{Arefiev2014}]. The dimensionless parameter $G$ is the ratio of the frequencies of oscillations induced by the field of the channel and by the field of the laser prior to a significant energy enhancement. Indeed, taking into account that the electron $\gamma$-factor in this regime is essentially $\gamma_{vac}$, we find that $\Omega / \omega \propto a_0 \omega_{p0} / \omega$ for $a_0 \gg 1$.  Currently, the dependence of $\gamma_{\max}$ on $G$ above the threshold remains a robust numerical observation that requires an in-depth theoretical analysis.

\begin{figure}
	\centering
	\includegraphics[width=0.95\columnwidth]{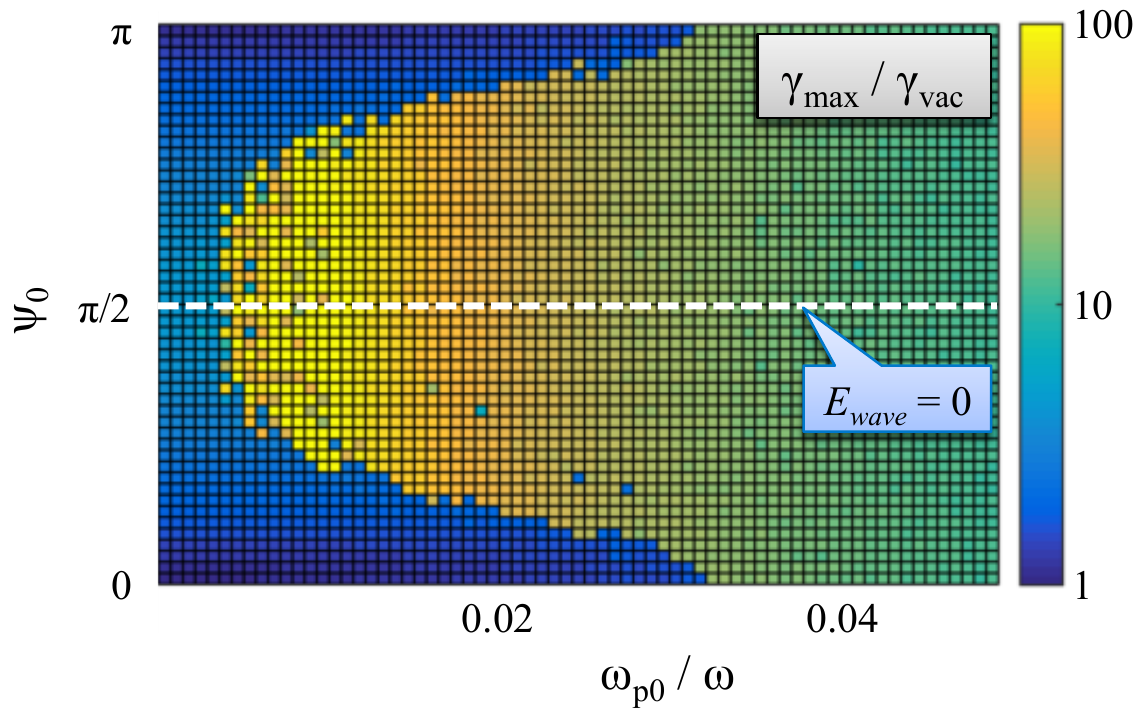} 
  \caption{The maximum $\gamma$-factor attained by injected electrons. The laser amplitude in all the cases is $a_0 = 8.5$. The initial injection phase defines the initial wave amplitude, $a = a_0 \sin(\psi_0)$.} \label{Figure_injection}
\end{figure}


\subsection*{Role of electron injection}

In the scan presented in Fig.~\ref{Invited_PoP_Figure_8} we have considered the case where initially the electron is already in the channel, but the laser pulse has not yet reached the electron location longitudinally. On the other hand, in the steady-state channel discussed in Sec.~\ref{Sec-channel} and illustrated in Fig.~\ref{fig:2D_example}, electrons are injected into the channel from the side with the laser beam already present in the channel. This raises the question regarding the role of electron injection in determining both the threshold and the energy gain. 

We test the role of injection by seeding an electron without any initial momentum into a plane wave with a given initial phase. Specifically, the wave amplitude is set to $a(\xi) = a_0 \sin(\psi_0-\xi)$, where $\psi_0$ is the phase of the laser pulse at the moment of injection defined as $\xi = 0$. The electron is placed onto the channel axis at the moment of injection. Figure~\ref{Figure_injection} shows how the maximum $\gamma$-factor attained by the electron depends on the injection phase $\psi_0$ and the ion density in the channel. We show only the range of initial phases $0 \leq \psi_0 \leq \pi$, because the plot for $\pi \leq \psi_0 \leq 2\pi$ is identical. 

The energy enhancement threshold is present for all the injection phases in Fig.~\ref{Figure_injection}, but the location of the threshold is sensitive to the injection phase. The threshold is the lowest when initially the electric field of the wave is equal to zero, $E_{wave} = 0$. Since the wave amplitude is fixed, this means that the injection phase determines the value of the parameter $G$ at the threshold. It should also be pointed that the energy gain increases as the threshold becomes lower. Therefore, we can conclude that proper electron injection into the laser beam can significantly enhance the electron energy gain from the wave, while lowering the corresponding threshold determined by the product $a_0 \omega_{p0} / \omega$.


\subsection*{Free oscillations across the channel}

\begin{figure}
	\centering
	\includegraphics[width=0.95\columnwidth]{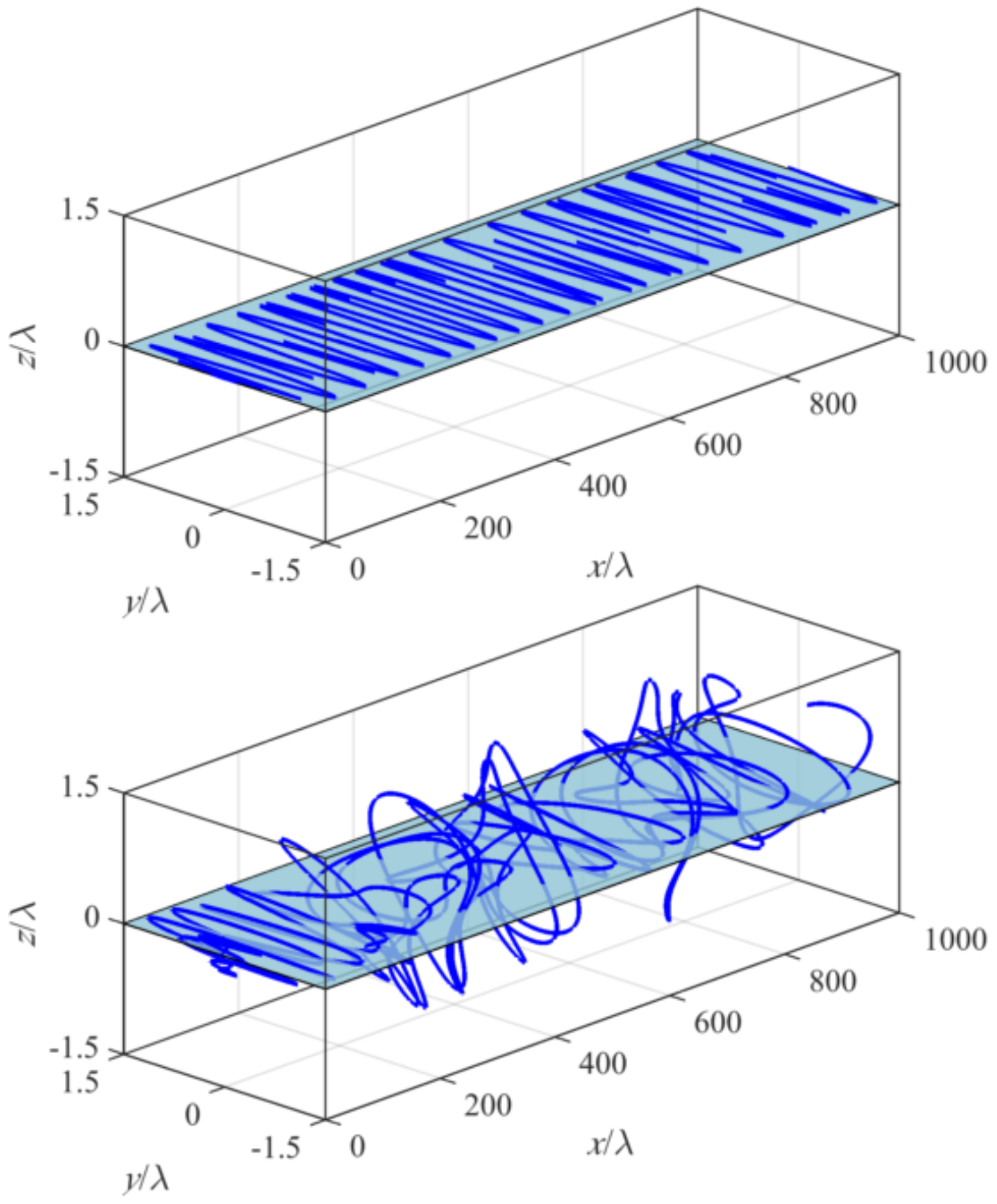} 
  \caption{Electron trajectories with and without an initial displacement (lower and upper panels) out of the plane of the driven oscillations. The wave electric field is polarized in the $(y,z)$-plane (shown in light-blue).} \label{Figure_9}
\end{figure}

We have so far deliberately limited the electron motion to the plane of the driven oscillations by placing the electron initially on the axis of the channel. Small off-axis  displacements in the absence of the laser pulse lead to small free oscillations across the channel since the restoring force is no longer equal zero. In what follows, we examine how the out-of-plane oscillations evolve in the presence of a laser pulse with relativistic amplitude $a_0 \gg 1$.   

The equation for the the out-of-plane oscillations can be directly obtain from Eq.~(\ref{perp_osc}) by taking into account that the channel is cylindrically symmetric. By setting the right-hand side to zero and replacing $y$ with $z$ we obtain an equation,
\begin{eqnarray} \label{free_perp_osc}
\frac{d^2 z}{d \tau^2} + \Omega^2 y = 0,
\end{eqnarray}
that resembles an equation for an oscillator with a natural frequency $\Omega$.

The free oscillations along the $z$-axis are coupled to the driven electron motion in the $(x,y)$-plane via the relativistic $\gamma$-factor, $\Omega \propto \sqrt{\gamma}$. As long as the amplitude of the free oscillations remains small, the $\gamma$-factor is determined predominantly by the driven motion. Taking the $\gamma$-factor in the vacuum regime shown in Fig.~\ref{Figure_DLA_vac_2} as an example of the driven motion, we can conclude that the driven motion affects the natural frequency $\Omega$ in two ways: it increases the natural frequency roughly by a factor of $a_0$ for $a_0 \gg 1$ and it also strongly modulates the natural frequency. The first aspect has been discussed in detail earlier in this Section and it is responsible for the onset of the significant energy enhancement. 

The modulation of the natural frequency can make the free oscillations along the $z$-axis parametrically unstable. The $\gamma$-factor modulations effectively modulate the restoring force acting on the oscillator described by Eq.~(\ref{free_perp_osc}). It is then intuitively clear that the oscillations remain stable as long as the modulation frequency is much higher than the natural frequency $\Omega$. The oscillations become unstable when the two frequencies become comparable. 

As already discussed, the driven electron motion below the energy enhancement threshold can be approximated by the solution for the vacuum regime. We again consider an electron that is irradiated by a plane wave in the channel and assume that the initial displacement is much smaller than $r_{\max}$. In this case, $\tau \approx -\xi/\omega$ and we then find from the $\gamma$ profile shown in Fig.~\ref{Figure_DLA_vac_2} that the modulation frequency is $\omega_{mod} \approx \omega$. On the other hand, significantly below the threshold we have $\Omega \ll \omega$, as evident from Fig.~\ref{Invited_PoP_Figure_8}. Therefore, free oscillations should be stable below the threshold.

As we cross the threshold by increasing the ion density, two important changes in the dependence of the $\gamma$-factor on $\tau$ take place. The $\gamma$-factor is significantly increased, which greatly increases the natural frequency. In addition to that, the modulations of the $\gamma$-factor become less frequent. This is because the peaks of enhanced $\gamma$ act as modulations and they take multiple laser cycles to develop. The combination of these two factors changes the relation between the two characteristic frequencies to $\Omega \gg \omega_{mod}$. We therefore conclude that the energy enhancement changes the stability of the free oscillations, making them susceptible to the parametric instability.

A parameter scan over the same range of $a_0$ and $\omega_{p0}/\omega$ as in Fig.~\ref{Invited_PoP_Figure_8} reveals that there is a threshold for the amplification of the free transverse oscillations~\cite{Arefiev2015b}. This threshold matches the energy enhancement threshold (dashed line in Fig.~\ref{Invited_PoP_Figure_8}), because the energy enhancement triggers the growth of the electron oscillations along the $z$-axis. The amplitude of the free oscillations grows until it becomes comparable to $r_{\max}$. 

The amplitude of the driven oscillations, i.e. the oscillations along the $y$-axis, also has an enhancement threshold that matches the energy enhancement threshold as well. We find that, in general, the electron energy enhancement is accompanied by a considerable enhancement of the transverse oscillations across the channel, making the electron trajectory considerably three-dimensional with the maximum displacement comparable to $r_{max}$. To illustrate this point, we show in Fig.~\ref{Figure_9} electron trajectories for $a_0 = 8.5$ and $\omega_{p0} / \omega \approx 0.24$. These parameters are above the threshold according to Fig.~\ref{Invited_PoP_Figure_8}. In the upper panel, the electron has no initial out-of-plane displacement, so its trajectory remains flat. However, an initial displacement of $\Delta z = 0.2 \lambda$ in the lower panel provides a necessary seed for the instability to develop significant out-of-plane oscillations. As a result, a non-planar electron trajectory quickly develops, accompanying the energy enhancement.


\section{Summary and discussion}

We have examined a regime in which a laser pulse with relativistic intensity irradiates a sub-critical plasma over a time period much longer than the characteristic electron response time. It is shown that a steady-state channel is formed in the plasma in this case with quasi-static transverse and longitudinal electric fields. These fields, even though they are much smaller than the electric field in the laser pulse, profoundly alter the electron dynamics. The longitudinal electric field reduces the longitudinal dephasing between the electron and the wave. This allows the electron to gain significantly more energy from the wave. The energy gain in this regime is ultimately limited by the superluminosity of the wave fronts induced by the plasma in the channel, even though the relativistic transparency greatly reduces the role of the plasma. The transverse electric field alters the oscillations of the transverse electron velocity. As a result, it can remain anti-parallel to the laser electric field, which leads to significant energy gain by the electron from the wave. We showed that this process has a sharp threshold determined by a single parameter, $a_0 \omega_{p0} / \omega$. The threshold can be greatly reduced and the energy gain further enhanced by appropriately injecting electrons into the laser pulse. The threshold for the energy enhancement matches the threshold for the onset of the parametric instability in the direction perpendicular to the plane of the driven electron oscillation. Therefore, the trajectories of electrons with enhanced energies eventually become non-planar if the interaction time and length with the pulse are sufficient for the instability to develop.

The effect of the longitudinal electric field is akin to pre-acceleration of electrons in the longitudinal direction. Analysis of pre-accelerated electrons in a channel~\cite{Arefiev2014} has showed that pre-acceleration reduces the energy enhancement threshold and increases the electron energy gain. This indicates that the transverse and longitudinal electric fields can synergistically enhance the electron  energy gain. Further in-depth research is required to determine the energy scalings and the key parameters that determine the electron dynamics. 

Finally, it should be pointed out that even though the presented analysis was motivated by laser-plasma interactions at significantly sub-critical plasma densities, most of the results are much more general and they are not limited just to the regime where $\omega_{pe} \ll \omega$. Therefore, the analysis is well suited to make meaningful predictions regarding electron acceleration in near-critical and over-critical plasmas, provided that such plasmas are relativistically transparent to the incoming high-intensity laser pulse~\cite{Lefebvre1995,Guerin1996}. However, wave propagation becomes a crucial aspect in this case and it should be addressed self-consistently~\cite{Toncian2016} taking into account transverse laser pulse dimensions and its polarization~\cite{Stark2015,Gonzalez2016}.

\section*{Acknowledgments}
AVA would like to thank Dr. Toma Toncian for constructive comments. This material is based upon work supported by the U.S. Department of Energy [National Nuclear Security Administration] under Award Number DE-NA0002723. A.V.A. was also supported by AFOSR Contract No. FA9550-14-1-0045, U.S. Department of Energy - National Nuclear Security Administration Cooperative Agreement No. DE-NA0002008, and U.S. Department of Energy Contract No. DE-FG02- 04ER54742. V.N.K. and G.S. were supported by AFOSR Contract No. FA9550-14-1-0045 and U.S. Department of Energy Contract Nos. DE-SC0007889 and DE-SC0010622. Simulations for this paper were performed using the EPOCH code (developed under UK EPSRC grants EP/G054940/1, EP/G055165/1 and EP/G056803/1) using HPC resources provided by the Texas Advanced Computing Center at The University of Texas. 


\appendix

\section{2D PIC simulation parameters} \label{Appendix_1}

The snapshots shown in Figs.~\ref{fig:2D_example} and \ref{Figure_DLA_Ez_2} are from two-dimensional particle-in-cell simulations performed using an open-source code EPOCH~\cite{Arber2015}. In both simulations, the initial electron density profile is given by
\begin{equation}
n_e =
   \begin{cases}
      n_0 \exp[-(x - x_0)^2/ L^2] ,  & \text{for $x <  x_0$;}\\
      n_0,  & \text{for $x \geq  x_0$},
   \end{cases} \label{n_sim}
\end{equation}
where $x_0 = 5$ $\mu$m and $L = 6$ $\mu$m. In Fig.~\ref{fig:2D_example}, we set $n_0 = 0.01 n_c$, whereas, in Fig.~\ref{Figure_DLA_Ez_2}, we set $n_0 = 0.05 n_c$. The electron population is initialized using 10 macro-particles per cell. The ion density is initially equal to the electron density and the ion density profile is initialized using 4 macro-particles per cell. The electrons are initially cold, whereas the ions are immobile throughout the simulation. The size of the domain in the simulation shown in Fig.~\ref{fig:2D_example} is 180 $\mu$m (12000 cells) along the $x$-axis and 60 $\mu$m (1200 cells) along the $y$-axis. The size of the domain in the simulation shown in Fig.~\ref{Figure_DLA_Ez_2} is 140 $\mu$m (9500 cells) along the $x$-axis and 60 $\mu$m (1200 cells) along the $y$-axis. The longitudinal resolution is chosen in agreement with the criterion outlined in Ref.~[\onlinecite{Arefiev2015c}] in order to correctly compute the electron energy gain.

We use a laser pulse whose focal plane in the absence of plasma is located at $x = 0$ $\mu$m. The laser fields have a Gaussian profile along the $y$-axis, with the electric and magnetic fields in the focal plane given by
\begin{equation}
\left. E_{wave} \right/ E_0 = \left. B_{wave} \right/ B_0 = S(t) \exp \left( - \left. y^2 \right/ w_0^2 \right),  
\end{equation}
where $w_0 = 8.5$ $\mu$m and the temporal profile is 
\begin{equation}
S(t) =
   \begin{cases}
      \exp[-(t - t_0)^2/ T^2] ,  & \text{for $t <  t_0$;}\\
      1,  & \text{for $t \geq  t_0$},
   \end{cases} \label{E_sim}
\end{equation}
with $t_0 = 182.4$ fs and $T = 85$ fs. In both simulations, the laser wavelength is 1 $\mu$m and the peak laser intensity at $x = y = 0$ $\mu$m is $I = 10^{20}$ W/cm$^2$, which corresponds to $a_0 = 8.5$. The normalizing amplitudes $E_0$ and $B_0$ are maximum electric and magnetic fields in a plane wave with $a_0 = 8.5$. In the simulation shown in Fig.~\ref{fig:2D_example}, the laser electric field has $x$ and $y$ components, whereas the laser magnetic field has only a $z$-component. In the simulation shown in Fig.~\ref{Figure_DLA_Ez_2}, the laser electric field has only a $z$-component, whereas the laser magnetic field has $x$ and $y$ components.



\end{document}